\documentclass{cs23proc}

\usepackage{kantlipsum}
\usepackage{dblfloatfix}

\editors{Takeru Suzuki and the Cool Stars 23 Organizing Team}
\publisher{Zenodo}
\conference{The 23th Cambridge Workshop on Cool Stars, Stellar Systems, and the Sun (Cool Stars 23)}
\conferencedate{2026}

\title{Systematic and Statistical Properties of Prominence Eruptions on the M-dwarf YZ CMi}
\author{
Yuto Kajikiya$^{1}$,
Kosuke Namekata$^{2,3,4,5}$,
Yuta Notsu$^{6,7}$,
Kai Ikuta$^{8}$,
Hiroyuki Maehara$^{9}$,
Bunei Sato$^{1}$,
Daisaku Nogami$^{5}$
}

\affiliation{
$^{1}$ Department of Earth and Planetary Sciences, Institute of Science Tokyo,
2-12-1 Ookayama, Meguro-ku, Tokyo 152-8551, Japan \\

$^{2}$ Heliophysics Science Division, NASA Goddard Space Flight Center,
8800 Greenbelt Road, Greenbelt, MD 20771, USA \\

$^{3}$ The Catholic University of America,
620 Michigan Avenue, N.E., Washington, DC 20064, USA \\

$^{4}$ The Hakubi Center for Advanced Research, Kyoto University,
Kyoto 606-8302, Japan \\

$^{5}$ Department of Physics, Kyoto University,
Kitashirakawa-Oiwake-cho, Sakyo-ku, Kyoto 606-8502, Japan \\

$^{6}$ Laboratory for Atmospheric and Space Physics, University of Colorado Boulder,
3665 Discovery Drive, Boulder, CO 80303, USA \\

$^{7}$ National Solar Observatory,
3665 Discovery Drive, Boulder, CO 80303, USA \\

$^{8}$ Graduate School of Social Data Science, Hitotsubashi University,
Kunitachi, Tokyo 186-8601, Japan \\

$^{9}$ Okayama Branch Office, National Astronomical Observatory of Japan (NINS),
3037-5 Honjo, Kamogata-cho, Asakuchi, Okayama 719-0232, Japan
}

\shorttitle{Prominence Eruptions on YZ CMi}
\shortauthors{KAJIKIYA ET AL.}
\abs{M-dwarfs produce frequent flares, and their associated mass ejections are expected to significantly affect the habitability of close-in exoplanets. Recent spectroscopic observations have revealed several prominence eruptions---indicative of stellar mass ejections---on M-dwarfs through Doppler shifts of the H$\alpha$ line. However, systematic and statistical studies, particularly regarding their association with white-light flares, have been limited due to the lack of intensive and continuous simultaneous photometric and spectroscopic monitoring of the same target star. We conducted one month of continuous spectroscopic observations of the active M-dwarf YZ CMi using the 3.8-m Seimei telescope with an unprecedentedly high time cadence of $\sim$1~min, simultaneously with TESS. We detected four prominence eruptions, among which two events showed rapid, short-duration eruptions with velocities of $\sim$300--500~km~s$^{-1}$ and durations of approximately 5~min. Such short-duration events may have been missed in previous observations due to insufficient time cadence. We further performed a systematic analysis using a total of 35 H$\alpha$ flares on YZ CMi observed simultaneously with TESS. Notably, most prominence eruptions (6 out of 7) were not associated with detectable white-light flares. This result suggests that most observed prominence eruptions on M-dwarfs may have occurred near the stellar limb, where white-light flares are difficult to detect, which may imply a potential observational bias in their detectability due to low contrast with the background emission. These first statistical constraints, together with the discovery of rapid, short-duration prominence eruptions, indicate that previous observations may have underestimated both the frequency and velocities of mass ejections on M-dwarfs due to observational biases and highlight the necessity of reassessing their impact on close-in exoplanets.}

\begin{document}

\maketitle

\section{Introduction}
M-dwarfs are the most abundant stars in the Galaxy and host several known Earth-sized planets within or near their habitable zones, making them among the most promising targets in the search for life beyond the Solar System \citep[e.g.,][]{Shields2016,Henry2024}. However, their frequent flares and associated coronal mass ejections (CMEs) may strongly influence planetary habitability. In particular, stellar mass ejections are theoretically expected to contribute to atmospheric escape, alter atmospheric chemistry, and modify the radiation and energetic-particle environments of planets located within the close-in habitable zones of M-dwarfs \citep[e.g.,][]{Airapetian2016,Yamashiki2019,Airapetian2020}. On the other hand, theoretical studies have suggested that the strong large-scale magnetic fields of active M-dwarfs may suppress mass ejections \citep[e.g.,][]{Alvarado-Gomez2018,Sun2022}. Observational constraints on physical properties of stellar mass ejections on M-dwarfs are therefore essential for evaluating the exoplanetary habitability.

Although direct detection of stellar CMEs remains difficult because stellar coronae cannot be spatially resolved, H$\alpha$ spectroscopy provides an indirect method for searching for prominence eruptions, which trace cool, dense plasma and may constitute the core of CMEs \citep[e.g.,][]{Otsu2022,Namekata2022a}. Erupting material moving toward the observer can produce transient Doppler-shifted components in the H$\alpha$ line profile. Such features allow us to investigate the line-of-sight velocity and temporal evolution of prominence eruptions, providing valuable insights into stellar mass ejections.

\begin{figure}[ht!]
	\centering
	\includegraphics[width=0.85\linewidth]{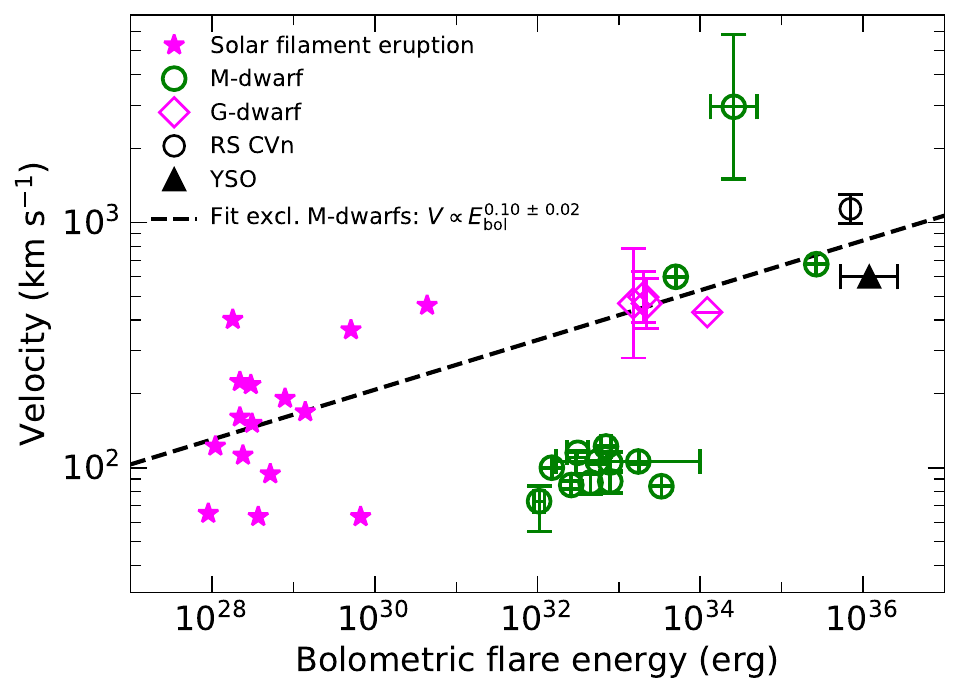}
    \caption{Velocities of stellar prominence-eruption candidates as a function of
    bolometric flare energy. The open green circles show the M-dwarf events
    reported in previous work \citep{Maehara2012,Moschou2019,Notsu2024}. The black filled upward triangle represents a blueshifted
    event on a young stellar object (YSO) reported in \citet{Moschou2019}. The
    pink open diamonds represent blueshifted events on the young Sun-like star
    EK Dra \citep{Namekata2022a,Namekata2024a,Namekata2025b}, while the black
    open circle denotes a blueshifted event on an RS CVn-type binary star
    \citep{Inoue2023}. Pink filled stars represent solar filament eruptions
    from \citet{Seki2019}. The dashed line represents the best-fitting power-law
    relation derived from the non-M-dwarf events.}
	\label{fig:Velocitu_M-dwarf}
\end{figure}

A growing number of prominence-eruption candidates have been detected on M-dwarfs \citep[e.g.,][]{Maehara2012,Vida2016,Vida2019,Inoue2024,Notsu2024}, raising further questions about the properties of stellar mass ejections. As shown in Figure~\ref{fig:Velocitu_M-dwarf}, most reported events on M-dwarfs have velocities of $\sim$100 km s$^{-1}$ \citep[e.g.,][]{Maehara2012,Inoue2024,Notsu2024}, which are lower than those observed on other types of stars \citep{Namekata2025b}, even at comparable flare energies. 
Moreover, the reported association rate of prominence-eruption candidates with flares having energies of $\gtrsim10^{32}$ erg, comparable to X10-class solar flares, is only $\sim$20$\%$ \citep{Notsu2024}, substantially lower than the flare--CME association rate for X-class solar flares ($\gtrsim90\%$; \citealt{Yashiro&Gopalswamy2009}). These relatively low velocities and occurrence rates may suggest magnetic suppression of mass ejections on M-dwarfs and, consequently, potentially  weaker impact on surrounding planetary environments than might be expected from their large flare energies.

However, current observational estimates may be affected by several observational biases. First, previous H$\alpha$ spectroscopic observations have been obtained with cadences of $\gtrsim$5min. Solar prominence eruptions often show velocity evolution on timescales shorter than 5 min \citep{Otsu2022}, and the stronger surface gravity of M-dwarfs could potentially lead to more rapid deceleration of erupting material. Consequently, observations with insufficient temporal resolution may fail to resolve short-lived, high-velocity components and may underestimate both the maximum velocities and occurrence rates of prominence eruptions.

Second, the detectability of prominence eruptions may potentially depend on flare properties and viewing geometry. Strong impulsive H$\alpha$ line broadening can obscure asymmetric components \citep{Liu2025}, whereas in weaker flares, Doppler-shifted components may be difficult to distinguish from the line-center enhancement. In addition, viewing geometry may affect the visibility of erupting material. In the case of the Sun, on-disk filament eruptions are observed in absorption, whereas off-limb prominences appear in emission. For M-dwarfs, however, theoretical calculations suggest that on-disk prominence eruptions may also appear in H$\alpha$ emission because of the relatively faint photospheric background \citep{Leitzinger2022}. If prominence eruptions can appear in emission both on the stellar disk and beyond the limb, their detectability may vary with projected position because of changes in the background radiation field. These potential observational biases may contribute to underestimates of both eruption velocities and occurrence rates.
%Second, the detectability of prominence eruptions may potentially depend on viewing geometry and radiative-transfer effects. On the Sun, eruptions projected against the stellar disk, commonly referred to as filament eruptions, are observed in absorption, whereas off-limb prominences are seen in emission. For M dwarfs, however, theoretical calculations suggest that erupting cool plasma projected against the stellar disk may also appear in H$\alpha$ emission because of the relatively faint photospheric background \citep{Leitzinger2022}. All reported prominence-eruption candidates on M dwarfs have been detected as blueshifted emission features \citep[e.g.,][]{Vida2019,Notsu2024}, while clear blueshifted absorption associated with eruptions has not been reported. If prominence eruptions can appear in emission both on the stellar disk and beyond the limb, their detectability may vary with projected position because the background radiation field changes from disk center to the limb and off-limb regions. This potential viewing-geometry bias may affect the inferred velocity distribution of prominence eruptions through projection effects.
%Quantifying this viewing-geometry bias is therefore important for interpreting the observed occurrence rates of stellar prominence eruptions.

To address these potential observational biases, continuous high-cadence spectroscopic observations of the same target star are essential for systematically investigating prominence eruptions and resolving their rapid velocity evolution. Simultaneous photometry is also valuable because white-light emission is thought to mainly trace impulsive energy deposition at flare footpoints \citep{Namekata2017,Watanabe2017,Namizaki2023}. In the case of the Sun, impulsive flares tend to show stronger white-light emission than long-duration events \citep{Watanabe2017}, while white-light signatures may be suppressed when flare footpoints are occulted near the stellar limb. Photometric monitoring could therefore provide information on both flare heating and viewing geometry. In addition, rotational brightness modulations can be used to infer starspot properties \citep{Ikuta2023,Mori2024}, providing further constraints on the possible locations of flare and prominence-eruption sites \citep{Namekata2024b}.

%White-light flare emission is generally associated with impulsive energy deposition in the lower stellar atmosphere, particularly near flare footpoints, and therefore provides information on the strength and impulsiveness of flare heating. Comparing prominence eruptions with white-light flare properties may thus help clarify whether the occurrence and kinematics of mass ejections are related to the energy deposition process during flares.
In this study, we performed simultaneous high-cadence spectroscopic and photometric observations of the active M-dwarf YZ CMi. Section~2 describes the observations, and Section~3 presents the data analysis and the identification of H$\alpha$ line-profile asymmetries. In Section~4, we present rapid, short-duration prominence eruptions, examine their masses and kinetic energies, and investigate the statistical relationships between H$\alpha$ line-profile asymmetries and flare properties using an expanded sample of 35 H$\alpha$ flares. Finally, we summarize our main results and implications in Section~5.

\section{Observations}

We observed YZ CMi using the KOOLS-IFU spectrograph mounted on the
3.8-m Seimei Telescope at Okayama Observatory, Japan
\citep{Kurita2020}. The observing campaign was conducted in 2021 January and February over a total of 13 clear nights, simultaneously with observations by the Transiting Exoplanet Survey Satellite (TESS; \citealt{2015JATIS...1a4003R}) during Sector~34. We used the VPH683 grism, which covers approximately
5800--8000~\AA\ with a spectral resolution of
$\lambda/\Delta\lambda \sim 2000$
\citep{Matsubayashi2025}. The exposure time was 60~s, and the detector
readout required approximately 17~s, resulting in a cadence of 77~s. This
cadence is substantially shorter than the typical duration of an
M-dwarf optical flare and enables us to resolve H$\alpha$ line-profile
variations on timescales of $\sim$1 minute.

\section{Analysis}
\subsection{Data Reduction and Flare Detection}
The spectra were reduced using standard procedures, including bias subtraction, flat-field correction, spectral extraction, wavelength calibration, and relative flux normalization, following the methods described in \citet{Namekata2020,Namizaki2023}. The reduction was performed using IRAF \citep{IRAF1986} and PyRAF \citep{Pyraf2012}. We measured the
H$\alpha$ equivalent width and inspected the temporal evolution of
the residual line profiles after subtracting a representative
quiescent spectrum.

We detected H$\alpha$ flares by requiring a consecutive enhancement exceeding $3\sigma$ above the quiescent pre-flare level \citep{Kajikiya2024}. Using this criterion, we identified 27 H$\alpha$ flares in the Seimei observations. The simultaneous TESS light curve was then used to determine whether each event was accompanied by a detectable white-light flare and to estimate its bolometric flare energy.

%For the statistical analysis of the white-light association, we
%combined the present events with previously reported simultaneous
%spectroscopic and TESS observations of YZ CMi \citep{Maehara2012,Notsu2024}. The combined
%sample contains 35 H$\alpha$ flares.

%The 27-event Seimei sample is used
%for the uniform line-profile classification described in
%Section~\ref{sec:classification}, %while the expanded 35-event sample
%is used to investigate the association with white-light emission.

\subsection{Identification of H$\alpha$ Line-profile Asymmetries}
\label{sec:classification}
Here, we describe the method used to identify H$\alpha$ line-profile asymmetries during H$\alpha$ flares. To avoid relying solely on visual inspection, we developed an automated procedure to classify the H$\alpha$ residual profiles as blue asymmetry, red asymmetry, or symmetry. The details of the method are presented in \citet{Kajikiya2024}.

First, we created the differential normalized spectra by subtracting the quiescent state pre-flare spectra from the flare state spectra (see Figure \ref{fig:rapid_eruption} (c)). 
Each flare-state residual profile was first fitted with a single
Voigt function representing the approximately symmetric flare
component. The same profile was then fitted with a composite model
consisting of a Voigt function and an additional Gaussian component (see Figure \ref{fig:rapid_eruption} (c)).
The additional Gaussian represents excess emission in either the blue
or red wing.
We compared the two models using the Bayesian Information Criterion
(BIC). Assuming Gaussian noise, the
BIC can be written as
\begin{equation}
\mathrm{BIC}
=
n\ln\left[
\frac{2\pi}{n}
\sum_{i=1}^{n}
\left\{F_i-f(\lambda_i)\right\}^2
\right]
+n+k\ln n ,
\end{equation}
where $F_i$ and $f(\lambda_i)$ are the observed and model fluxes,
respectively, $n$ is the number of spectral data points, and $k$ is
the number of free parameters in the model. We then defined
\begin{equation}
\Delta\mathrm{BIC}
=
\mathrm{BIC}_{\rm Voigt}
-
\mathrm{BIC}_{\rm Voigt+Gaussian}.
\end{equation}

A profile was classified as asymmetry when $\Delta{\rm BIC}>2$ in at least two consecutive spectra. The
profile was classified as blue or red asymmetry according to whether
the centroid of the Gaussian component was located on the blue or red
side of the adopted H$\alpha$ line center. Profiles that did not
satisfy these conditions were classified as symmetry.
Applying this procedure to the 27 H$\alpha$ flares, we identified
three events showing significant blue asymmetries, five showing
significant red asymmetries, and 19 without significant asymmetries.
%The automated classification reduces subjective differences between
%events and permits a uniform comparison of the line profiles and
%their white-light counterparts.

\section{Results and Discussions}

\subsection{Rapid, Short-duration Prominence Eruptions \citep{Kajikiya2024}}
We identified four prominence eruptions among the three blue-asymmetry and five red-asymmetry events, based on the velocities and temporal evolution of the asymmetric H$\alpha$ components. We do not discuss the detailed classification criteria here; see \citet{Kajikiya2024} for a more detailed discussion and individual case studies. The primary reason for interpreting these events as prominence eruptions is the large velocities of the asymmetric components ($\sim$300--500~km~s$^{-1}$; \citealt{Kajikiya2024}). These velocities are substantially higher than those expected from alternative processes that can produce blueshifted or redshifted H$\alpha$ components, such as chromospheric evaporation and condensation \citep[e.g.,][]{Allred2006, Tei2018}. In addition, the observed velocities, durations, and temporal evolution are broadly consistent with those of solar prominence eruptions.

%Several other phenomena can also produce red- and blueshifted H$\alpha$ components, such as chromospheric evaporation, chromospheric condensation, and post-flare loops. Nevertheless, we interpret these four events as prominence eruptions. We do not discuss the detailed classification here; see \citet{Kajikiya2025a} for a more detailed discussion.

\begin{figure*}[!t]
    \centering
    \includegraphics[width=0.75\linewidth]{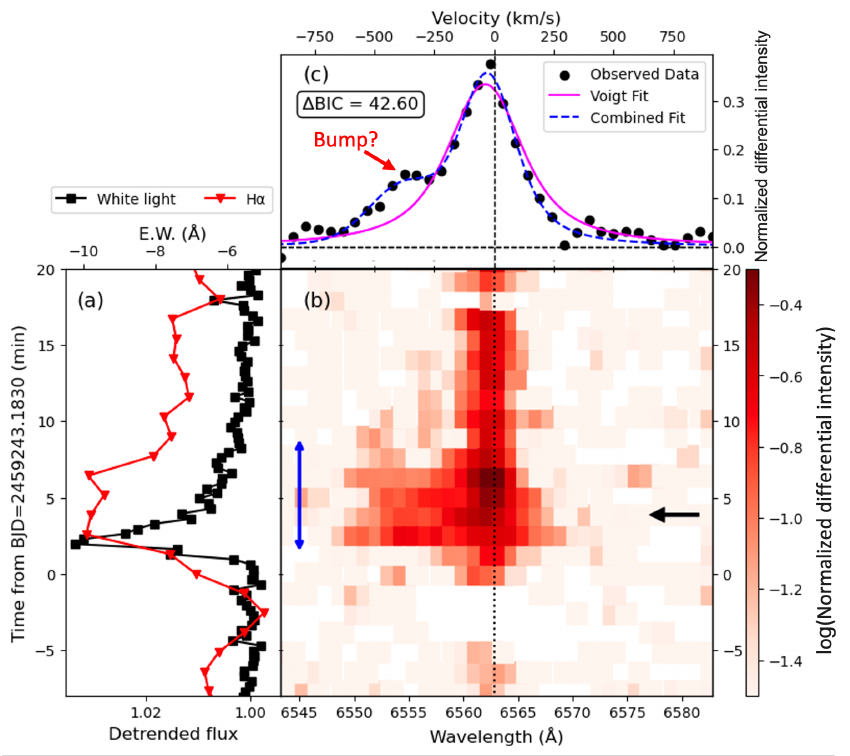}

\caption{Rapid, short-duration prominence-eruption candidate in flare Y8.
(a) White-light (black) and H$\alpha$ equivalent-width (red) light curves.
(b) Temporal evolution of the H$\alpha$ differential line profile. The
horizontal axis shows wavelength, and the vertical axis shows elapsed time
from flare onset. The color bar shows the logarithm of the normalized
differential intensity. The black arrow marks the time of maximum
$\Delta\mathrm{BIC}$, and the blue bidirectional arrow indicates the duration
of the blue asymmetry. (c) Differential H$\alpha$ spectrum at the time of
maximum $\Delta\mathrm{BIC}$. The top axis shows Doppler velocity relative to
6562.8~\AA. Black dots show the observed spectrum, the magenta solid line
shows the Voigt fit, and the blue dashed line shows the combined
Voigt+Gaussian fit.}
    \label{fig:rapid_eruption}
\end{figure*}

\begin{figure*}[!t]
    \centering
    \includegraphics[width=0.75\linewidth]{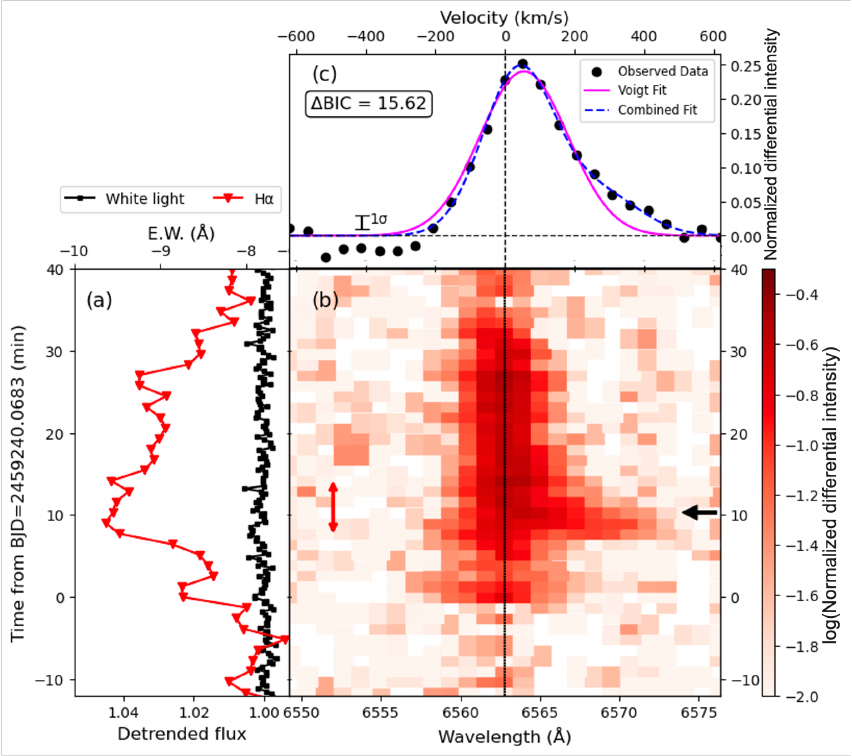}
\caption{Same as Figure~2, but for the rapid, short-duration backward
prominence-eruption candidate in flare Y6. The red bidirectional arrow in
panel (b) indicates the duration of the red asymmetry.}
    \label{fig:rapid_backeruption}
\end{figure*}

%The primary reason for this interpretation is the large Doppler velocities of the blue- and redshifted components. The observed line-of-sight velocities are typically $\sim$200--500~km~s$^{-1}$ \citep{Kajikiya2025a}. Although cool upflows associated with chromospheric evaporation can produce blueshifted H$\alpha$ emission, the observed velocities are substantially higher than those typically reported in solar observations \citep{Svestka1962,Tei2018,Huang2019,Li2019} and predicted by radiative-hydrodynamic models of M-dwarf flares \citep{Allred2006}. In addition, the observed velocities, durations, and temporal evolution are broadly consistent with those of solar prominence eruptions. We therefore interpret these four events as prominence-eruption candidates.
Here, we present two representative prominence eruptions identified in this study. Figures~\ref{fig:rapid_eruption} and \ref{fig:rapid_backeruption} show the light curves and temporal evolution of the H$\alpha$ line profiles (dynamic spectra) of these events. The inferred velocities reach approximately 300--500~km~s$^{-1}$, while the asymmetric components persist for only $\sim$5~min, as shown in Figures~\ref{fig:rapid_eruption}(b) and \ref{fig:rapid_backeruption}(b). The dynamic spectra further show rapid velocity evolution on timescales of $\sim$1~min, indicating that the eruptive motions developed and evolved on very short timescales. These events are among the shortest-duration red- and blue asymmetry events reported for stellar flares and could be identified owing to the high time cadence of our observations.
Interestingly, the red asymmetry event shows no clear white-light flare, although a marginal enhancement may be present (see Figure~\ref{fig:rapid_backeruption}(a)). We interpret this event as a backward prominence eruption occurring near the stellar limb, where the flare footpoints responsible for the white-light emission are hidden behind the stellar disk. This geometry can explain the absence of a detectable white-light enhancement despite the presence of a strong redshifted H$\alpha$ component.

We note that three of the four prominence eruptions detected in this study show rapid velocity changes on timescales of $\sim$1 min and short durations of only $\sim$5--10 min, substantially shorter than those of previously reported events. This suggests that such short-duration events and rapid velocity evolution may be common on M-dwarfs and may have been missed in earlier observations because of insufficient time cadence. Therefore, previous observations may have substantially underestimated the velocities and occurrence rates of prominence eruptions, highlighting the importance of high-cadence observations for studying mass ejections on M-dwarfs.

%Two of the prominence-eruption candidates, one redshifted event (Y6) and one blueshifted event (Y8), exhibited particularly high velocities of approximately 300--500~km~s$^{-1}$ and very short durations of only $\sim$5~min. These events are among the shortest-duration red- and blueshifted asymmetry events reported for stellar flares and could be identified owing to the high time cadence of our observations. 
%Figure~\ref{fig:rapid_eruption} shows the light curve and temporal evolution of the H$\alpha$ line profiles for one of these rapid events, Y8. The residual H$\alpha$ spectra, obtained by subtracting the quiescent spectrum, show a clear blueshifted emission component (Figure~\ref{fig:rapid_eruption}(c)). This blueshifted component evolved rapidly and disappeared within approximately 8~min.

%With a cadence of
%5--10~min, the high-velocity phase could be sampled only once or missed
%entirely. Temporal averaging would additionally dilute the wing
%emission and reduce the apparent maximum velocity. Therefore, the
%observed distribution of stellar prominence-eruption velocities may
%be biased toward slower and longer-lived events.

\subsection{Bolometric Energy vs. Kinetic Energy}
We estimated the masses and kinetic energies of the four prominence eruptions, as well as the bolometric energies of the associated flares, following previous studies \citep[e.g.,][]{Maehara2021,Inoue2024}. The derived properties are summarized in Table \ref{tab:eruption_properties}. For events with non-white-light flares, we estimated the bolometric flare energy using the empirical scaling relation between the bolometric flare energy and H$\alpha$ flare energy \citep{Namekata2024a}.

\begin{figure*}[!t]
    \centering
    \includegraphics[width=1\textwidth]{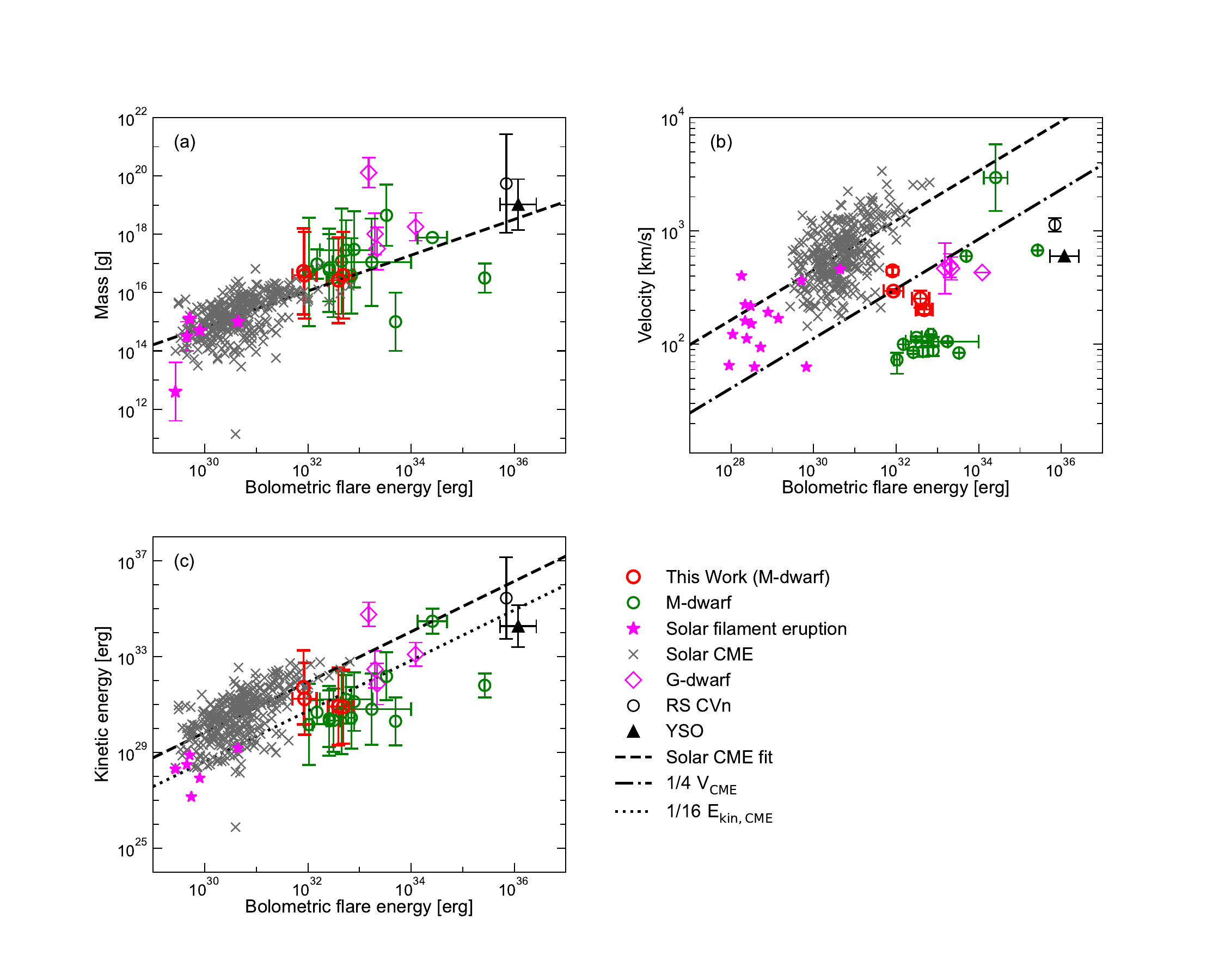}
    \caption{Mass, velocity, and kinetic energy of solar and stellar prominence
eruptions and CMEs as functions of flare bolometric energy. (a) Mass of
prominence eruptions and CMEs as a function of flare bolometric energy. Red
open circles show the four prominence-eruption candidates reported in this
study \citep{Kajikiya2024}. Green open circles represent blue asymmetry
events on M-dwarfs reported in \citet{Maehara2012,Moschou2019,Notsu2024}. The black filled upward triangle represents a
blue asymmetry event on a young stellar object (YSO) reported in \citet{Moschou2019}. Pink open diamonds represent blue asymmetry events on the young
Sun-like star EK Dra \citep{Namekata2022a,Namekata2024a,Namekata2025b},
while the black open circle denotes a blue asymmetry event on an RS CVn-type
binary star \citep{Inoue2023}. Pink filled stars represent solar filament
eruptions and surges from \citet{Namekata2022a}, while gray crosses represent
solar CMEs from \citet{Yashiro&Gopalswamy2009}. The dashed black line shows the
power-law fit to the solar CME data shown by the gray crosses. (b) Velocity of
prominence eruptions and CMEs as a function of flare bolometric energy. Pink
filled stars represent solar filament eruptions from \citet{Seki2019}; the
other symbols are the same as in panel (a). The dash-dotted black line shows
one-quarter of the fitted solar CME velocity relation. (c) Kinetic energy of
prominence eruptions and CMEs as a function of flare bolometric energy.
Symbols are the same as in panel (a). The dotted black line shows one-sixteenth
of the fitted solar CME kinetic-energy relation.}

    \label{fig:Kin_vs_Bol}
\end{figure*}

Figure \ref{fig:Kin_vs_Bol} shows the comparison of the masses, velocities, and kinetic energies of our events with those of previously reported solar and stellar prominence eruptions and mass-ejection events (e.g., \citealt{Namekata2024b,Notsu2024}).  As shown in Figure \ref{fig:Kin_vs_Bol}(a), the estimated prominence masses approximately follow the scaling relation derived for solar CMEs and are consistent with previously reported events on M-dwarfs \citep{Maehara2012,Notsu2024}. 
Although the velocities of our events are substantially higher than those of previously reported M-dwarf prominence eruptions observed with lower time cadence, they still lie below the scaling relation for solar CMEs. This difference may arise because the observed velocities correspond to prominence eruptions rather than to the CME itself, as also discussed by \citet{Maehara2012,Notsu2024}. In the solar case, CME velocities are typically approximately four times higher than those of the associated prominence eruptions \citep{Gopalswamy2003}. Indeed, our events are approximately consistent with a velocity scaling relation reduced to one quarter of the solar CME relation. Correspondingly, because the kinetic energy scales as $E_{\rm kin}\propto v^2$, the events are also broadly consistent with a kinetic-energy relation reduced to $1/16$ of the solar CME scaling, as shown in Figure \ref{fig:Kin_vs_Bol}(b).
Although the current sample is too small for a robust statistical conclusion, our results suggest that high-time-cadence spectroscopy may recover faster prominence eruptions that could have been missed or underestimated in previous lower-cadence observations of M-dwarfs. Further high-cadence spectroscopic observations are therefore essential.

%compaird the relation between bolometric flare energy and kinetic energy of prominence eruptions infered from observation, by plot on previous solar and stellar prominence eruption and mass ejection events (e.g., \citep{Namekata2024b,Notsu2024}). 
%The measured velocities are projected line-of-sight values and do not
%directly provide the three-dimensional velocities of the ejecta.
%Nevertheless, the detection of transient components reaching several
%hundred kilometres per second demonstrates that at least some
%M-dwarf eruptions evolve on substantially shorter timescales than
%previously resolved.

\begin{table*}[t]
    \centering
    \caption{
    Physical properties of the four prominence-eruption candidates detected on YZ~CMi in this work. 
    }
    \label{tab:eruption_properties}
    \begin{tabular*}{0.96\textwidth}
    {@{\extracolsep{\fill}} l c c c c}
    \hline\hline
    Event &
    $E_{\rm bol}$ &
    $v_{\rm max}$ &
    $M_{\rm p}$ &
    $E_{\rm kin}$ \\
     &
    ($10^{32}$ erg) &
    (km s$^{-1}$) &
    ($10^{16}$ g) &
    ($10^{31}$ erg) \\
    \hline
    Y6 &
    $0.86^{+0.62}_{-0.36}$ &
    $295\pm12$ &
    $3.9^{+116}_{-3.8}$ &
    $1.7^{+53.3}_{-1.7}$ \\

    Y8 &
    0.82 &
    $445\pm36$ &
    $5.4^{+155}_{-5.2}$ &
    $5.2^{+175}_{-5.0}$ \\

    Y12 &
    $3.89^{+2.35}_{-1.46}$ &
    $254\pm43$ &
    $2.6^{+75.4}_{-2.6}$ &
    $0.84^{+34.2}_{-0.82}$ \\

    Y20 &
    $4.81^{+2.81}_{-1.78}$ &
    $202\pm18$ &
    $3.9^{+116}_{-3.8}$ &
    $0.80^{+27.2}_{-0.78}$ \\
    \hline
    \end{tabular*}
\end{table*}

\subsection{Systematic and Statistical Properties of H$\alpha$ Line-profile Asymmetries \citep{Kajikiya2025a}}
We investigated the systematic and statistical relationship between H$\alpha$ line-profile asymmetries and flare properties. Our sample consists of 35 H$\alpha$ flares on YZ~CMi observed simultaneously with TESS, including 27 flares analyzed in this work and eight flares reported in previous studies \citep{Maehara2012,Notsu2024}. Among these events, seven were interpreted as prominence eruptions. This relatively large and homogeneous sample, obtained for the same target and with the same spectral diagnostic, enables a systematic statistical investigation of H$\alpha$ line-profile asymmetries. Here, we focus particularly on the dependence of prominence-eruption occurrence on flare energy and on their association with white-light flares. For the statistical properties related to stellar rotational phase and spot distribution, we refer the reader to \citet{Kajikiya2025a}.
%In solar flares, the CME association rate is known to increase with flare energy \citep{Yashiro&Gopalswamy2009}. In addition, white-light flares are generally associated with intense energy deposition in the lower atmosphere and tend to occur in regions with strong magnetic fields around the flare energy-release sites \citep{Watanabe2017}. These solar trends motivate us to investigate whether the occurrence and properties of prominence eruptions on YZ~CMi also depend on flare energy and white-light emission.

\begin{figure*}[t]
    \centering
    \includegraphics[width=0.45\linewidth]{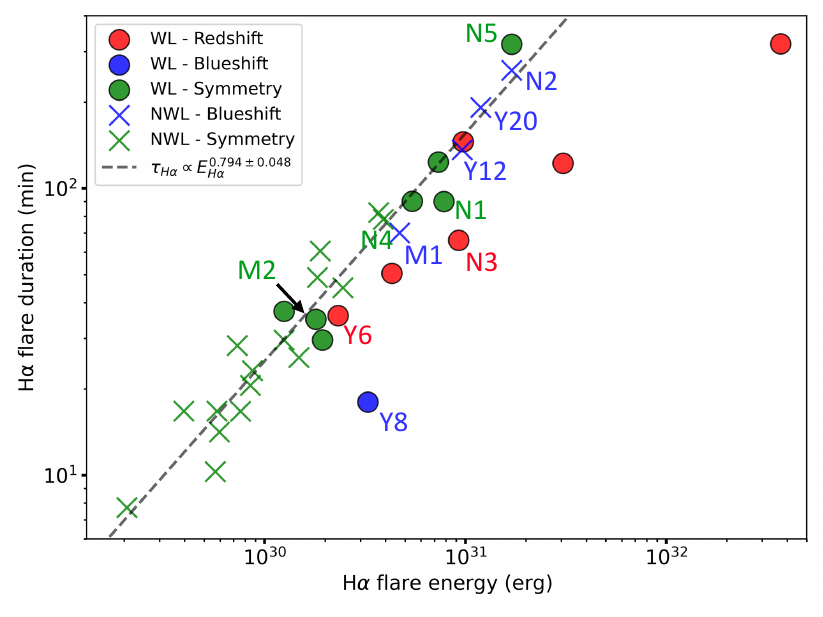}
    \caption{
    Relation between H$\alpha$ flare energies and durations. The difference in color represents the asymmetry type, while the difference in marker style represents whether the flares are white-light flares or non-white-light flares. The black dashed line indicates the fitted line for only symmetry events.
    }
    \label{fig:Energy_statistics}
\end{figure*}

First, we compared the properties of the asymmetry events with the H$\alpha $ flare energies and durations. Figure~\ref{fig:Energy_statistics} shows the relation between H$\alpha$ flare energy and duration, color-coded by the type of line-profile asymmetry. As shown in this figure, higher-energy flares tend to show a higher occurrence rate of both red- and blue asymmetries. We note all H$\alpha$ flares showing significant line-profile asymmetry have energies above $2\times10^{30}$~erg. Although the number of blue asymmetry events interpreted as prominence eruptions is still limited, this trend is qualitatively consistent with the increasing CME association rate toward higher-energy solar flares \citep{Yashiro&Gopalswamy2009}. However, this apparent lower-energy threshold could reflect an observational selection effect. In low-energy flares, the flare-related H$\alpha$ emission is weaker, potentially making asymmetric components more difficult to detect. Above $2\times10^{30}$~erg, both red- and blue asymmetry events span a wide range of flare energies and durations. %Nevertheless, the occurrence rate of line-profile asymmetries shows an increasing trend with flare energy above this threshold.

%Among the eight flares with energies above $9\times10^{30}$~erg, seven show significant asymmetry, whereas only four out of ten flares in the energy range of $2$--$9\times10^{30}$~erg show asymmetry. These results suggest that line-profile asymmetries occur more frequently even in higher-energy flares.

%\subsubsection{Line profile asymmetry vs. white-light flare assocuation}

\begin{figure*}[t]
    \centering
    \includegraphics[width=0.75\linewidth]{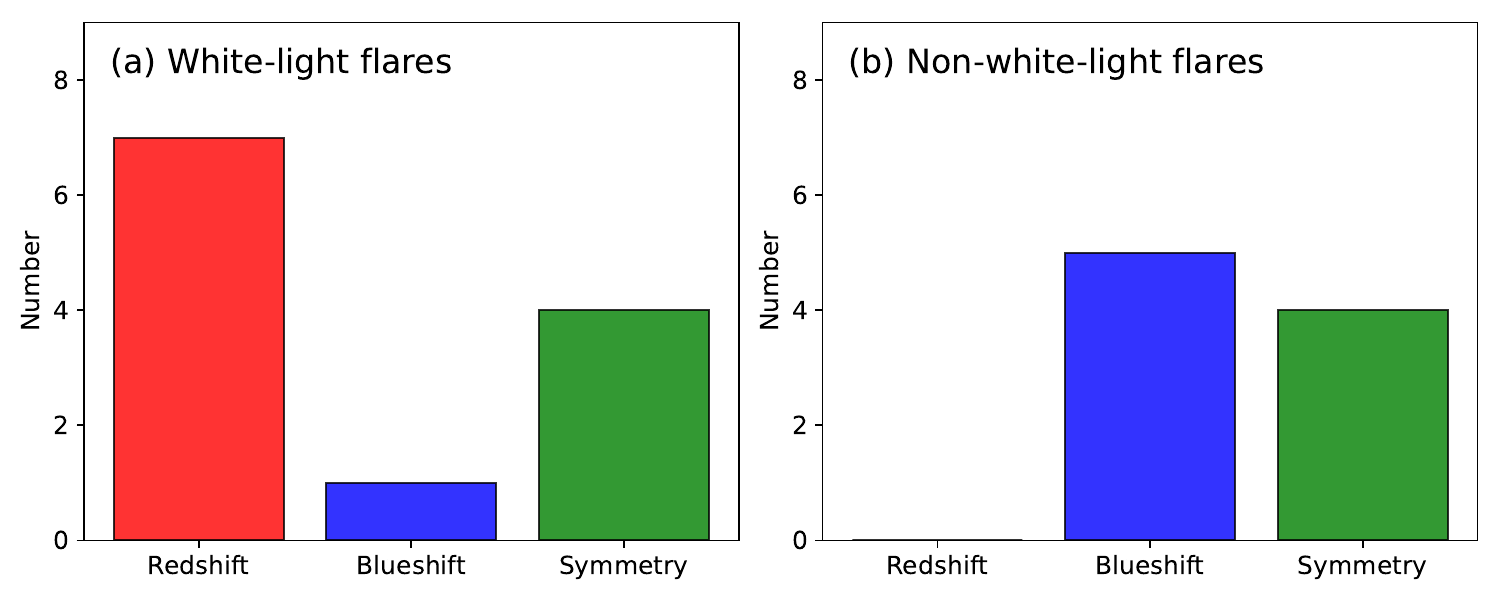}
    \caption{
    Comparison of histograms for (a) white-light flares and (b) non-white-light flares categorized by asymmetry type. The vertical axis represents the number of asymmetry events (Events with $E_{\mathrm{H\alpha}} > 2 \times 10^{30}$ erg). We note that if one flare exhibits multiple independent asymmetries, each asymmetry is counted separately.
    }
    \label{fig:WL_statistics}
\end{figure*}

Next, we investigated the dependence of H$\alpha$ line-profile asymmetries on the association with white-light flares. Figures~\ref{fig:WL_statistics}(a) and (b) show the white-light association of the H$\alpha$ flare events. As shown in these figures, all seven events showing red asymmetries are associated with white-light flares. This result suggests that the origin of the red asymmetry may be related to the energy deposition associated with white-light emission. This tendency is consistent with an interpretation of chromospheric condensation \citep{Namizaki2023}, which is expected to accompany strong impulsive heating and can contribute to the formation of white-light flare emission \citep{Kowalski2018,Kowalski2024}. In contrast, five out of the six blue-asymmetry events, which are interpreted as prominence eruptions, are not associated with white-light flares. This is notable because white-light flares are more numerous than non-white-light flares in the sample shown in Figure~\ref{fig:WL_statistics}. 
One possible explanation is that these flares have relatively low heating rates and therefore do not produce detectable white-light emission, similar to non-white-light flares observed on the Sun \citep{Watanabe2017}.

Another possible explanation is a viewing-geometry effect. If a flare occurs near the stellar limb, white-light emission associated with the flare footpoints may be partially or completely occulted by the stellar disk, while erupting prominence material extending off the limb may remain visible. The visibility of H$\alpha$ prominence emission may also vary depending on the projected position \citep{Leitzinger2022}. Because the optical continuum around H$\alpha$ is relatively faint in M-dwarfs, the contrast of prominence emission against the stellar disk may differ from that in the Sun and solar-type stars and may also vary between on-disk and off-limb geometries \citep{Leitzinger2022}. Such differences in the visibility of white-light emission and H$\alpha$ emission from prominences could potentially contribute to the lack of detectable white-light counterparts for some prominence eruptions. If off-limb events are more readily detected, projection effects may also contribute to relatively low measured line-of-sight velocities. We note that this interpretation remains qualitative, and further radiative-transfer modeling and observational studies are needed to test this possibility.
%Consequently, the kinetic energies inferred directly from these velocities could be underestimated as well. The apparent lack of white-light emission in most of the prominence-eruption candidates may potentially indicate that H$\alpha$ prominence signatures and white-light flare emission have different visibility functions across the stellar disk.

\section{Summary and Conclusions}
In this study, we conducted high-time-cadence spectroscopic observations of the active M-dwarf YZ~CMi, simultaneously with TESS. We detected 27 H$\alpha$ flares, among which five showed red asymmetries and three showed blue asymmetries in their H$\alpha$ line profiles. We identified four of these events as prominence eruptions, including three blue asymmetry events and one red asymmetry event. We also introduced a new methodology based on the Bayesian Information Criterion (BIC) to automatically identify asymmetric components in H$\alpha$ line profiles.

In particular, we detected two rapid, short-duration prominence eruptions with velocities of $\sim$300--500~km~s$^{-1}$ and durations of only $\sim$5~min. These events are among the shortest-duration stellar prominence eruptions reported so far. Such rapid, short-duration events can be missed or temporally diluted in observations with cadences longer than several minutes. Our detections therefore suggest that previous lower-cadence observations ($\gtrsim5$~min) may have underestimated both the velocities and occurrence rates of prominence eruptions on M-dwarfs.

We also investigated the systematic and statistical properties of H$\alpha$ line-profile asymmetries during flares. We found that higher-energy flares tend to show red- or blue asymmetries more frequently. In addition, the occurrence of asymmetry shows a clear dependence on the association with white-light flares. All seven red asymmetry events in the combined sample are associated with white-light flares, which is consistent with an interpretation of chromospheric condensation. In contrast, five out of the six blue asymmetry events, interpreted as prominence eruptions, are not associated with white-light flares. We note the redshifted backward prominence eruption shows only a marginal white-light enhancement, although it satisfies our white-light-flare detection criterion.
The weak white-light association may partly reflect a viewing-geometry bias favoring the detection of off-limb prominence eruptions, which could also lead to underestimated line-of-sight velocities.

Overall, our results suggest that previous H$\alpha$ spectroscopic observations may have substantially underestimated both the occurrence rate and characteristic velocities of mass ejections on M-dwarfs because of limited time cadence and potential viewing-geometry biases. Further high-time-cadence observations will be essential for determining the intrinsic occurrence rates, velocities, and kinetic energies of stellar mass ejections, and for reassessing their potential impact on the atmospheres and habitability of close-in planets.

\section*{Acknowledgments}
{The authors thank the organizers of Cool Stars 23 for the opportunity to present this work in a plenary oral session. This research was supported by JSPS (Japan Society for the Promotion of Science) KAKENHI Grant Numbers  JP21J00316 (K.N.), JP20K04032, JP24K00685 (H.M.), JP24K17082 (K.I.), JP24K00680 (K.N., H.M., and D.N.), and JP24H00248 (Y.K., K.N., K.I., H.M., and D.N). Y.N. acknowledges the funding support from NASA ADAP 80NSSC21K0632, and NASA TESS Cycle 6 80NSSC24K0493.
The spectroscopic data used in this paper were obtained through the program 21A-N-CN03 (PI: K.N.) with the 3.8m Seimei telescope, which is located at Okayama Observatory of Kyoto University.
This paper includes data collected with the TESS mission, obtained from the MAST data archive at the Space Telescope Science Institute (STScI). Funding for the TESS mission is provided by the NASA Explorer Program.
STScI is operated by the Association of Universities for Research in Astronomy, Inc., under NASA contract NAS 5-26555. 
Some of the data presented in this paper were obtained from the Mikulski Archive for Space Telescopes (MAST) at the Space Telescope Science Institute. The specific observations analyzed can be accessed via https://doi.org/10.17909/14ym-zt14. The authors acknowledge ideas from the participants in the workshop ``Blazing Paths to Observing Stellar and Exoplanet Particle Environments" organized by the W.M. Keck Institute for Space Studies.}

\bibliographystyle{cs23proc}
\bibliography{example.bib}

\end{document}